\documentstyle[aps,preprint]{revtex}
\begin{document}
\draft
\title{Conservation of the spectral moments \\
in the n-pole approximation}
\author{F. Mancini\footnote{Corresponding author.  E-mail address: mancini@vaxsa.csied.unisa.it}}
\address{Dipartimento di Scienze Fisiche ``E.R. Caianiello" e Unit\`a I.N.F.M. di Salerno \\
Universit\`a di Salerno, 84081 Baronissi (SA), Italy}
\maketitle

\begin{abstract}
A formulation of the Green's function method is presented in the  $n$-pole 
approximation. Without referring to a specific model we give a general scheme of 
calculations that easily permits the computation of the ``single-particle" Green's 
function in terms of the energy matrix. A theorem is proved which states that the 
moments of the spectral density function are conserved up to the order $2(n-l+1)$, 
where $l$ is the order of the composite field. A comparison with the spectral density 
approach is also discussed.
\end{abstract}


	The theoretical analysis of highly correlated electron systems contains many unsolved 
problems and a complete and systematic formulation is still lacking. Many analytical methods 
have been developed. Among many we recall the slave boson method [1], the non-crossing 
approximation [2], the $d_\infty$ method [3], the projection operator method [4], the method of 
equation of motion [5], the spectral density approach [6], the coherent potential 
approximation [7], the composite operator method [8]. It is not easy to judge the reliability of 
the various approximations. Certainly, one important aspect is the capability to conserve the 
symmetries inherent to the model and some relations, which can be derived on general basis.

	A scheme of approximation, common to various methods [4-6, 8], is the so called  $n$-pole 
approximation, based on a linearization of the equations of motion. In this article we 
show that in this scheme the exact relation between the spectral moments and the spectral 
density is conserved: {\it (i)} at any order when the moments are calculated with respect to the
linearized equations: {\it (ii)} up to the order $2(n-l+1)$, where $l$ is the order of the 
composite field, when calculated with respect to the full equations of motion. 
This aspect is attractive and shows that the approximation can reproduce some 
essential features of the single-particle function.

	Let us consider a certain Hamiltonian 
\begin{equation}
H=H(\phi_1,\ldots ,\phi_p)
\end{equation}      
the set $\{\phi_i\}$ denotes field operators. In order to close the infinite hierarchy of equations of 
motion some truncation is necessary. One procedure is to choose a basis of operators 
$\{\psi_1, \ldots, \psi_n\}$ and linearize the equations of motion as
\begin{equation}
i{\partial\over \partial t}\psi(x) = [\psi(x), H]=\epsilon (-i\vec\nabla)\psi (x) + \delta j (x) \approx
\epsilon (-i\vec\nabla)\psi(x)
\end{equation}	                           
where the eigenvalue or energy matrix $\epsilon$ [9] is self-consistently calculated by means of the 
equation
\begin{equation}
\epsilon(-i\nabla_x)\langle\left[ \psi({\rm\bf x}, t),\psi^\dagger ({\rm\bf y}, t)\right]_\pm\rangle =
\langle \left[\left[ \psi({\rm\bf x}, t),H\right]\psi^\dagger ({\rm\bf y}, t)\right]_\pm\rangle
\end{equation}	                
The symbol $\left[\ldots \right]_\pm$ denotes equal-time anticommutator or commutator, in dependence of the 
statistics of the set $\{\psi_i\}$. The rank of the energy matrix is equal to $n$, the number of 
components of the vector $\psi(x)$. When the basis $\{\psi_i\}$ coincides with the original set $\{\phi_i\}$, 
the linearization procedure given in Eq. (2) corresponds to the Hartree-Fock approximation.
The fields $\psi_i$ may be called composite fields [8], since they are generally constructed
as a combination of the original fields $\phi_i$.

	Let us consider the thermal retarded Green's function [10]
\begin{equation}
S(x,y)=\langle R\left[\psi(x) \psi^\dagger (y) \right] \rangle = {i\over (2\pi)^{d+1}} \int d^d kd\omega e^{i{\rm\bf k}\cdot
({\rm\bf x}\cdot{\rm\bf y})-i\omega (t_x-t_y)} S({\rm\bf k},\omega)
\end{equation}  	                        
where R is the usual retarded operator and the symbol $\langle\ldots\rangle$ denotes the thermal average. 
 $d$ is the dimensionality of the system. By means of the Heisenberg equation (2), in the linearized form,
 the Fourier transform $S({\rm\bf k},\omega)$ is given by
\begin{equation}
S({\rm\bf k},\omega)={1\over \omega-\epsilon({\rm\bf k})} I({\rm\bf k}) = {1\over D({\rm\bf k}, \omega) }\sum^{n-1}_{i=0}	
\omega^i\lambda^{(i)} ({\rm\bf k})
\end{equation}           
where 
\begin{equation}
I({\rm\bf k}) = F.T. \langle \left[ \psi ({\rm\bf x},t) ,\psi^\dagger ({\rm\bf y},t)\right]_\pm\rangle
\end{equation}	         
the symbol $F.T.$ denoting the Fourier transform, and
\begin{equation}
D({\rm\bf k},\omega)=\sum_{i=0}^n \omega^ia_i ({\rm\bf k})
\end{equation}	     
is the characteristic polynomial of the matrix $\epsilon({\rm\bf k})$. The characteristic coefficients $a_i({\rm\bf k})$ are 
defined by the following relation
\begin{equation}
a_{n-k}({\rm\bf k}) = (-1)^k Tr_k [\epsilon({\rm\bf k})]\qquad \hbox{for}\quad 0\le k\le n
\end{equation}	                 
Where $Tr_k[\epsilon]$ is the trace of the $k$'th order, defined as the sum of the determinants of all  
$\left({n\atop k}\right)$ matrices of order $(k\times k)$ that can be formed by intersecting any $k$ 
rows of $\epsilon$ with the same $k$ columns. We note that $Tr_n[\epsilon]=Det[\epsilon]$ and the convention 
$Tr_0[\epsilon]=1$ is used. The $\lambda^{(i)}({\rm\bf k})$ are the $n\times n$ matrices, determined by the 
equation
\begin{equation}
\sum^{n-1}_{i=0} \omega^i \lambda^{(i)} ({\rm\bf k}) =D ({\rm\bf k}, \omega)
\left[ \omega-\epsilon({\rm\bf k})\right]^{-1} I({\rm\bf k}) 
\end{equation}           
The solution of this equation gives:
\begin{equation}
\lambda^{(m)}({\rm\bf k}) = \sum_{s=m+1}^n a_s ({\rm\bf k}) \epsilon^{s-m-1} ({\rm\bf k})	I({\rm\bf k})\qquad \hbox{for}
\quad 0\le m\le n-1
\end{equation}                    
The Green's function (5) has a  $n$-pole structure
\begin{equation}
S({\rm\bf k},\omega) = \sum_{i=1}^n{\sigma^{(i)} ({\rm\bf k})\over \omega - E_i({\rm\bf k} +i\eta}
\end{equation}	             
The energy spectra $E_i({\rm\bf k})$ are the characteristic values of the matrix $\epsilon({\rm\bf k})$ and correspond 
to the zeros of the characteristic polynomial
\begin{equation}
\sum_{m=0}^n a_m ({\rm\bf k})E^m_i({\rm\bf k}) = 0
\end{equation}	       
The spectral functions are given by 
\begin{equation}
\sigma^{(i)}({\rm\bf k}) = {1\over b_i({\rm\bf k})} \sum_{m=0}^{n-1} E^m_i ({\rm\bf k}) \lambda^{(m)} ({\rm\bf k})
\end{equation}	          
where we put 
\begin{equation}
b_i({\rm\bf k}) = \prod_{j=1, j\neq i}^n \left[ E_i ({\rm\bf k}) - E_j({\rm\bf k})\right]
\end{equation}	        
We also note that in the $n$-pole expansion, the Green's function can be expressed as 
\begin{equation}
S_{\alpha\beta} ({\rm\bf k}, \omega) = {I_{\alpha\beta}({\rm\bf k}) \over \omega - 
\Sigma_{\alpha\beta} ({\rm\bf k}, \omega)}
\end{equation}
where the self energy $\Sigma_{\alpha\beta}({\rm\bf k}, \omega)$ is given by
\begin{equation}
\Sigma_{\alpha\beta}({\rm\bf k}, \omega) ={1\over {\displaystyle\sum_{i=0}^{n-1}} \omega^i 
\lambda^{(i)}_{\alpha\beta}({\rm\bf k})} \sum_{i=0}^{n-1} \omega^i[\epsilon ({\rm\bf k}) \lambda^{(i)} 
({\rm\bf k})]_{\alpha\beta}
\end{equation} 

	Summarizing, in this scheme of approximation the calculation of the Green's function 
requires the following steps:

\noindent {\it (i)} Given a certain Hamiltonian, a basic set $\{\psi_i\}$ of field operators is chosen;

\noindent{\it (ii)} the energy matrix $\epsilon({\rm\bf k})$ is calculated by means of Eq. (3);

\noindent{\it (iii)} the energy spectra $E_i({\rm\bf k})$ are calculated as the roots of the characteristic polynomial of 
$\epsilon({\rm\bf k})$;

\noindent{\it (iv)} the $\lambda^{(i)}({\rm\bf k})$ matrices are calculated in terms of the energy matrix $\epsilon({\rm\bf k})$
 by means of Eq. (10).

	We want now to study if the $n$-pole approximation conserves the exact relation between 
the spectral moments and the spectral density, that can be derived on general basis. We start 
by the following definition for the spectral moments
\begin{equation}
M^{(k)} ({\rm\bf k}) = F.T. \left[ \left( {i\partial \over \partial t_x}\right)^{k-p} 
\left(- {i\partial \over \partial t_y}\right)^{p} \langle \left[ \psi(x) ,\psi^\dagger (y)\right]_\pm \rangle \right]
_{t_x=t_y} \quad [\hbox{for}\ 0\le p\le k] 	          
\end{equation}        
We note that $M^{(k)} ({\rm\bf k})$ is a $n\times n$ matrix and is a generalization of the quantity usually 
introduced in the literature, where the spectral moments are referred to the one-particle 
propagator. Let us introduce the spectral density function   $A({\rm\bf k},\omega)$   
\begin{equation}
A({\rm\bf k},\omega) = -{1\over \pi} Im S ({\rm\bf k}, \omega +i\eta)
\end{equation}	        
and consider the quantity 
\begin{equation}
B^{(k)} ({\rm\bf k}) = \int^{+\infty}_{-\infty} d\omega \omega^k A({\rm\bf k},\omega)
\end{equation}	       
By taking time derivatives of Eq. (4) we find the following relation between the spectral 
moments and the spectral density
\begin{equation}
M^{(k)} ({\rm\bf k})=B^{(k)} ({\rm\bf k})
\end{equation}	     
This relation is an exact one and must hold for any integer non negative value of $k$. By means 
of the field equation Eq. (2)  the spectral moments can be written as
\begin{equation}
M^{(k)} ({\rm\bf k}) =\tilde M^{(k)} ({\rm\bf k}) +\delta M^{(k)} ({\rm\bf k})  
\end{equation}          
where
\begin{equation}
\tilde M^{(k)} ({\rm\bf k})	=\epsilon^k ({\rm\bf k})I({\rm\bf k})
\end{equation}       
\begin{equation}
\delta M^{(0)} ({\rm\bf k}) = \delta M^{(1)} ({\rm\bf k}) =0
\end{equation}
\begin{equation}
\delta M^{(k)} ({\rm\bf k}) = \sum_{m=1}^{k-1}	F.T. \left[ \epsilon^{k-m-1} (i\partial /\partial t_x)^m
\langle\left[\delta j(x), \psi^\dagger (y)\right]_\pm\rangle \right]_{t_x=t_y} \qquad
[\hbox{for}\quad k\ge 2]
\end{equation}                 
What see that the $\tilde M^{(k)} ({\rm\bf k})$ correspond to the spectral moments calculated in the $n$-pole
approximation, where the higher order field $\delta j(x)$ in the Heisenberg equation (2) is neglected.
We will now prove that the spectral moments $\tilde M^{(k)} ({\rm\bf k})$ satisfy the relation
\begin{equation}
\tilde M^{(k)}({\rm\bf k}) = \int^{+\infty}_{-\infty} d\omega\omega^k A({\rm\bf k},\omega)
\qquad \hbox{for any}\quad k 	
\end{equation}           
The proof is divided in two steps:
(i)  $\tilde M^{(k)}({\rm\bf k})$ and $B^{(k)}({\rm\bf k})$  satisfy for $k\ge n$ the same recurrence relation;
(ii) the equality $\tilde M^{(k)}({\rm\bf k})= B^{(k)}({\rm\bf k})$ is satisfied for $k\le n$.

	By recalling that $a_s({\rm\bf k})$ are the coefficients of the characteristic polynomial of the matrix 
$\epsilon({\rm\bf k})$, the Hamilton-Cayley theorem says 
\begin{equation}
\sum_{s=0}^n a_s({\rm\bf k})\epsilon^s({\rm\bf k}) =0
\end{equation}	          
Then, the moments $\tilde M^{(k)}({\rm\bf k})$ satisfy the recurrence relation
\begin{equation}
\tilde M^{n+k}({\rm\bf k})  = -\sum^{n-1}_{s=0} a_s ({\rm\bf k}) \tilde M^{s+k}({\rm\bf k})\qquad \hbox{for}\quad k\ge 0
\end{equation}            
On the other hand, by noting that from Eq. (11) the spectral density is given by 
 $A({\rm\bf k},\omega) = {\displaystyle \sum^n_{i=1}} \sigma^{(i)} ({\rm\bf k}) \delta\left[ \omega - E_i({\rm\bf k}) \right]$, 
we have
\begin{equation}
B^{(k)} ({\rm\bf k}) = \sum^n_{i=1} E^k_i ({\rm\bf k}) \sigma^{(i)} ({\rm\bf k})
\end{equation}	           
Since the $E_i({\rm\bf k})$ satisfy the relation
\begin{equation}
E_i^{n+k} ({\rm\bf k}) = -\sum^{n-1}_{m=0} a_m ({\rm\bf k}) E^{m+k}_i ({\rm\bf k})\qquad\hbox{for}
\quad k\ge 0
\end{equation}	                  
the quantities $B^{(k)}({\rm\bf k})$ satisfy the recurrence relation
\begin{equation}
B^{(n+k)} ({\rm\bf k}) = -\sum^{n-1}_{m=0} a_m ({\rm\bf k}) B^{(m+k)} ({\rm\bf k})\qquad\hbox{for}
\quad k\ge 0
\end{equation}	                  
We see that the spectral moments $\tilde M^{(k)} ({\rm\bf k})$ and the quantities $B^{(k)} ({\rm\bf k})$ 
satisfy the same recurrence relation for $k\ge n$. To show that they coincide, it is enough to 
consider the case $0\le k\le n$.

	Let us take $1\le k\le n-1$. By means of the expression (13) for the spectral functions we 
can write
\begin{equation}
B^{(k)} ({\rm\bf k}) = \sum_{m=0}^{n-1} \lambda^{(m)} ({\rm\bf k}) P^{(k+m)} ({\rm\bf k})
\end{equation}	        
where we have defined 
\begin{equation}	
P^{(k)} ({\rm\bf k}) \equiv \sum^n_{i=1} {E^k_i({\rm\bf k})\over b_i({\rm\bf k})}
\end{equation}	       
By means of the relation (29) the quantities $P^{(k})({\rm\bf k})$ satisfy the recurrence relation
\begin{equation}
\sum^n_{m=0} a_m ({\rm\bf k}) P^{m+k} ({\rm\bf k}) =0\qquad \hbox{for}\quad k\ge 0
\end{equation}	             
By simple algebraic relations it is easy to show that 
\begin{equation}
P^{(k)}({\rm\bf k})=\Biggl\{\begin{array}{ll}
0 \quad & \hbox{for}\quad 0\le k\le n-2\\
1 \quad & \hbox{for}\quad k= n-1
\end{array}
\end{equation}        
By recalling that from Eq. (10) $\lambda^{(m)} = \epsilon\lambda^{(m+1)} + a_{m+1} I$
we can then write
\begin{eqnarray}
B^{(k)} ({\rm\bf k})  &=& \lambda^{(n-1)} ({\rm\bf k}) P^{(k+n-1)} ({\rm\bf k}) + \epsilon ({\rm\bf k}) 
\sum_{m=n-1-k}^{n-2} \lambda^{(m+1)} ({\rm\bf k}) P^{(k+m)} ({\rm\bf k}) \\
&& + I({\rm\bf k})\sum_{m=n-1-k}^{n-2} a_{m+1} ({\rm\bf k}) P^{(k+m)} ({\rm\bf k})
\end{eqnarray}	               
On the other hand  $\lambda^{(n-1)} ({\rm\bf k})=a_nI({\rm\bf k})$, therefore the previous expression takes the form
\begin{equation}
B^{(k)} ({\rm\bf k}) =\epsilon ({\rm\bf k})  B^{(k-1)} ({\rm\bf k}) +I({\rm\bf k}) \sum _{m=n-1-k} ^{n-1}
a_{m+1} ({\rm\bf k}) P^{(k+m)} ({\rm\bf k})  = \epsilon ({\rm\bf k})  B^{(k-1)} ({\rm\bf k})
\end{equation}               
where the relation (33) for $k\ge 1$ has been used. But for $k=0$ it is easy to see that 
$ B^{(0)} ({\rm\bf k})=I({\rm\bf k})$, therefore
\begin{equation}
 B^{(k)} ({\rm\bf k}) = \epsilon^k  ({\rm\bf k})I ({\rm\bf k}) = \tilde M^{(k)} ({\rm\bf k}) \qquad \hbox{for}\quad 
0\le k\le n-1
\end{equation}	           
Because of the recurrence relations (27) and (30), the equality (38) holds also for $k=n$ and the step 
(ii) is thus proved. Then, the equality  $\tilde M^{(k)} ({\rm\bf k}) = B^{(k)} ({\rm\bf k}) $ is true for any 
$k$ and the relation (25) is demonstrated.

We now turn to the more general question if the $n$-pole approximation conserves the full 
spectral moments $M^{(k)}({\rm\bf k})$, calculated by the exact Heisenberg equations. 
In the context of the Hubbard model it is known that in the 2-pole approximation the first 
4 spectral moments relative to the electronic field are conserved [6,11]. For higher order
moments the situation is not known.

It is always possible to choose the basic set as
\begin{equation}
i{\partial\over \partial t} \psi_l (x) = [\psi_l(x), H] = \sum_{p=1}^{l+1} \gamma _{lp} (-i\nabla) \psi_p (x)
\qquad [\hbox{for } 1\le l\le n-1]
\end{equation}
For this choice we have that 
\begin{equation}
\begin{array}{l}
\epsilon_{lp} =\gamma_lp\\ \delta j_l=0
\end{array}
\qquad \hbox{for } 1\le l,p\le n-1
\end{equation}
and the matrix $\epsilon$ has the following structure
\begin{equation}
\epsilon_{lp}= 0\quad \hbox{for} \ \Biggl\{ \begin{array}{l}
l\le n-2\\ p\ge l+2
\end{array}
\end{equation}
From the definition (24) and by use of the exact Heisenberg equation (38) we can derive 
the following recurrence relation
\begin{equation}
\delta M^{(k)}_{lp} ({\rm\bf k}) = \epsilon_{lr} ({\rm\bf k}) \delta M^{(k-2)}_{rs} ({\rm\bf k}) 
\epsilon^\dagger_{sp} ({\rm\bf k})\qquad \hbox{for}\   1\le l,p\le n-1
\end{equation}
Then, by means of Eqs. (24) and (40) we find that the matrix $\delta M^{(k)} ({\rm\bf k})$
has the following property
\begin{equation}
\delta M^{(2k)}_{lp} ({\rm\bf k}) = \delta M^{(2k+1)}_{lp} ({\rm\bf k}) = 0 \qquad \hbox{for}\   1\le l,p\le n-k
\end{equation}
By recalling Eq. (25) we can then enunciate the following theorem.

Given a set of fields $\{\psi_l, l=1, \ldots, n\}$, if the subset $\{\psi_l, l=1, \ldots, n-1\}$ satisfies linear Heisenberg equations 
$i\left(\partial/\partial t\right)\psi_l(x) = {\displaystyle \sum^{l+1}_{p=1}} \gamma_{lp} (-i\nabla)\psi_p(x)$, 
then the first $2(n-l+1$) spectral moments for the fields $\psi_l$ $[1\le l\le n-1]$ are conserved [12].

The theorem proved in this article shows that the $n$-pole approximation is equivalent to 
the SDA [6], when the polar ansatz is considered. The following remark is worth to be 
mentioned. In the SDA the unknown quantities, the energy spectra $E_i({\rm\bf k})$ and the spectral 
functions $\sigma^{(i)}({\rm\bf k})$, are evaluated in terms of the spectral moments through the equation 
\begin{equation}
M^{(k)} ({\rm\bf k}) = \sum_{i=1}^n E_i^k ({\rm\bf k}) \sigma^{(i)}({\rm\bf k})
\end{equation} 	            
To calculate one matrix element $S_{\alpha\beta} ({\rm\bf k},\omega)$ of the Green's function one must solve a set 
of $2n$ non-linear equations. When $n$ is not small this procedure is not very convenient. 
Indeed, most of the calculations in the SDA have been restricted to the case of two poles. In 
the $n$-pole approximation we only need to find the roots of the characteristic polynomial of 
the energy matrix. Once the energy spectra $E_i({\rm\bf k})$ are calculated, the $n\times n$ 
Green's function can be immediately computed, as we have previously shown.  

	In conclusion, in this paper we have presented a general formulation based on the  $n$-pole 
expansion for the calculation of the ``single-particle" Green's function. The denomination 
``single-particle" is used in a generalized sense, and refers to the propagator of the  
$n$-component composite field $\{\psi_i\}$. Without recurring to a specific model, 
we have presented a scheme of calculations which allows us to determine the Green's function 
once the energy matrix  $\epsilon({\rm\bf k})$ is known. This is the fundamental quantity and 
contains the dynamics, controlled by the Hamiltonian, the algebra, determined by the 
statistics of the original fields $\{\phi_i\}$, the boundary conditions, since it refers to expectation values.
The conservation of the spectral moments has been analyzed. We have shown that there is an 
internal consistency in the $n$-pole approximation, in the sense that the exact relation
between the spectral density function and the spectral moments, when calculated on the same ground,
 is preserved at any order. We have demonstrated a theorem that states that the full spectral moments,
calculated on the basis of the exact
Heisenberg equations, are conserved up to a certain order that depends on the number of poles.
For example, in tha case of a single-band electron model, the spectral moments of the electronic
field are conserved up to the $2n$th order. 
 The method is equivalent to the spectral density approach. One advantage of this formulation 
with respect to the SDA is that we only need to solve the characteristic equation of 
$\epsilon({\rm\bf k})$. 

\acknowledgements
The author  thanks D. Villani for a critical reading of the manuscript. Valuable discussions with A. Avella  
and M. Marinaro are gratefully acknowledged.


\begin{references}

\item	S.E. Barnes, J. Phys. F {\bf 6}, 1375 (1976); ibid {\bf 7}, 2637 (1977); 
P. Coleman, Phys. Rev. B {\bf 29}, 3035 (1984); G. Kotliar and A.E. Ruckenstein, Phys. Rev. 
Lett. {\bf 57}, 1362 (1986).
\item	Y. Kuramoto, Z. Phys. B {\bf 53}, 37 (1983); N. Grewe, Z. Phys. B {\bf 53}, 271 (1983); 
Th. Pruschke, Z. Phys. B {\bf 81}, 319 (1990).
\item	W. Metzner and D. Vollhardt, Phys. Rev. Lett. {\bf 62}, 324 (1989); A. Georges and G. 
Kotliar, Phys. Rev. B {\bf 45}, 6479 (1992); A. Georges and W. Krauth, Phys. Rev. B {\bf 48}, 
7167 (1993); M.J. Rosenberg, G. Kotliar and X.Y. Zhang, Phys. Rev. B B {\bf 49}, 10191 
(1994).
\item	K.W. Becker, W. Brenig and P. Fulde, Z. Phys. B B {\bf 81}, 165 (1990); H. Mori, Progr. 
Theor. Phys. B {\bf 33}, 423 (1965); ibid B {\bf 34}, 399 (1965); A.J. Fedro, Yu Zhou, T.C. Leung, 
B.N. Harmon and S.K. Sinha, Phys. Rev. B B {\bf 46}, 14785 (1992); P. Fulde, Electron 
Correlations in Molecules and Solids (Springer-Verlag, Berlin-Heidelberg, 1993); N.M. 
Plakida, V.Yu. Yushankhai and I.V. Stasyuk, Physica C {\bf 162-164}, 787 (1989); B. 
Mehlig, H. Eskes, R. Hayn and M.B.J. Meinders, Phys. Rev. B {\bf 52}, 2463 (1995).
\item	D.J. Rowe, Rev. Mod. Phys. {\bf 40}, 153 (1968); L.M. Roth, Phys. Rev. {\bf 184}, 451 (1969); 
J. Beenen and D.M. Edwards, Phys. Rev. B {\bf 52}, 13636 (1995).
\item	O.K. Kalashnikov and E.S. Fradkin, Sov. Phys. JETP {\bf 28}, 317 (1969); Phys. Stat. Sol. 
(b) {\bf 59}, 9 (1973); W. Nolting, Z. Phys. {\bf 255}, 25 (1972); G. Geipel and W. Nolting, 
Phys. Rev. B {\bf 38}, 2608 (1988); W. Nolting and W. Borgel, Phys. Rev. B {\bf 39}, 6962 
(1989); A. Lonke, J. Math. Phys. {\bf 12}, 2422 (1971); H. Eskes, A.M. Oles, M.B.J. 
Meinders and W. Stephan, Phys. Rev. B {\bf 50}, 17980 (1994); T. Schneider, M.H. 
Pedersen and J.J. Rodriguez-N\`unez, Z. Phys. B {\bf 100}, 263 (1996); T. Herrmann and W. 
Nolting, cond-mat/9702022.
\item	B. Velicky, S. Kirkpatrick and H. Erhenreich, Phys. Rev. {\bf 175}, 747 (1968); P. Soven, 
Phys. Rev. {\bf 178}, 1136 (1969).
\item	F. Mancini, S. Marra, A.M. Allega and H. Matsumoto,  Physica C {\bf 235-240}, 2253 
(1994); F. Mancini, S. Marra and H. Matsumoto, Physica C {\bf 244}, 49 (1995); {\bf 250}, 184 
(1995);  {\bf 252}, 361 (1995).
\item	Derivative operators $\lambda(-i\vec\nabla)$ are defined as $\lambda (-i\vec\nabla)
f({\rm\bf x}) =\int d^dy\lambda ({\rm\bf x},{\rm\bf y} f({\rm\bf y})$.
\item	For the sake of simplicity we are assuming translational invariance.
\item A. Avella, F. Mancini, D. Villani, L. Siurakshina and V. Yu Yushankhai, {\it The Hubbard model 
in the two-pole approximation}, cond-mat/9708009 Int. Journ. Mod. Phys. B (in print).
\item	For example, let us consider the Hubbard model described by the Hamiltonian 
$H=\sum_{ij} c^\dagger (i) c(j) + U\sum_i n_\uparrow (i) n_\downarrow(i)$, where 
$c(i)$ is the electron field and we are using 
the standard notation. In the 3-pole approximation the choice $\psi(i) = \left(\begin{array}{c}
c(i)\\ \eta(i) \\ \pi(i) \end{array} \right)$ for the basic set, where 
$\eta(i)-n(i)c(i)$ and $\pi(i)=[\eta(i), H] - U\eta(i)$, $n(i)$ being the density 
operator, will ensure that the first 6 moments relative to the electron field $c(i)$ and the first 
4 moments relative to the Hubbard field $\eta(i)$ are fully conserved.
\end{references}
\end{document}